\documentclass[aip,apl,reprint,english,twocolumn,superscriptaddress,longbibliography]{revtex4-2}
\usepackage{amssymb}
\usepackage{amsmath}
\usepackage{graphicx}
\usepackage{natbib}
\usepackage{epstopdf}
\usepackage{color}
\usepackage{braket}
\usepackage{physics}
\usepackage{mathrsfs}
\usepackage{upgreek}
\usepackage{todonotes}
\usepackage[colorlinks=true, allcolors=blue]{hyperref}

\begin{document}
\title{Spectroscopy of low-lying valley states in hot Si/SiGe quantum dots}

\author{Connor Nasseraddin}
\affiliation{Department of Physics and Astronomy, University of California, Los Angeles, California 90095, USA}
\affiliation{Center for Quantum Science and Engineering, University of California, Los Angeles, California 90095, USA}
\affiliation{Department of Electrical and Computer Engineering, University of California, Los Angeles, California 90095, USA}

\author{Heun Mo Yoo}
\affiliation{Department of Physics and Astronomy, University of California, Los Angeles, California 90095, USA}

\author{Tanner M. Janda}
\affiliation{Department of Physics and Astronomy, University of California, Los Angeles, California 90095, USA}
\affiliation{Center for Quantum Science and Engineering, University of California, Los Angeles, California 90095, USA}

\author{Jason R. Petta}
\altaffiliation{Author to whom correspondence should be addressed: petta@physics.ucla.edu}
\affiliation{Department of Physics and Astronomy, University of California, Los Angeles, California 90095, USA}
\affiliation{Center for Quantum Science and Engineering, University of California, Los Angeles, California 90095, USA}
\affiliation{Department of Electrical and Computer Engineering, University of California, Los Angeles, California 90095, USA}

\begin{abstract}
The presence of low-lying valley states in Si may hinder the development of large-scale spin-based quantum processors. Rapid prototyping of novel Si/SiGe heterostructures and gate stacks will be central to identifying pathways that increase the valley splitting. We compare the performance of pulsed-gate spectroscopy (PGS) and detuning axis spectroscopy (DAPS) at temperatures up to 700 mK. 
We find that DAPS outperforms PGS, with DAPS resolving valley splittings as small as $\sim$86 $\mu$eV, while the energy resolution of PGS is only $\sim$210~$\mu$eV. Our work demonstrates that DAPS can be used to efficiently extract valley splittings at elevated temperatures in high throughput cryostats. 
\end{abstract}

\maketitle

Spin qubits \cite{Loss1998,Burkard2023RMP} fabricated using Si/SiGe heterostructures have demonstrated high fidelity one \cite{wu2025simultaneoushighfidelitysinglequbitgates} and two qubit operations, \cite{xue2022quantum,mills2022twoqubit,Noiri2022} long range coupling of qubits, \cite{Borjans2020,Dijkema2025} coherent transport of spins over relatively large length scales,\cite{seidler2022,DeSmet2025,undseth2026weightfourparitycheckssilicon} the fabrication of larger quantum dot (QD) arrays, \cite{Neyens2024,tidjani2025threedimensionalarrayquantumdots} and initial demonstrations of quantum error correction. \cite{takeda2022,HRL2026} While Si is a desirable material due to its long spin coherence time \cite{Tyryshkin2012} and compatibility with conventional semiconductor fabrication processes, \cite{Ha2022,George2024} the presence of low-lying valley states degrades qubit readout, results in spin relaxation hot spots, and adversely impacts gate fidelities. \cite{ando_electronic_1982, mills2022high,Philips2022, yang2013, Volmer2025} Heterostructures incorporating thin Si quantum wells (QWs) and sharp interfaces are commonly used to increase the valley splitting, \cite{chen_detuning_2021} but statistically relevant instances of small valley splitting threaten the development of large scale Si-based quantum processors. Alternative methods to increase the valley splitting have been proposed, such as introducing a constant \cite{Losert2023} or modulated \cite{mcjunkin2021valley} Ge concentration in the Si QW, but these approaches have not been tested at scale.

\begin{figure}[tbh!]
	\centering
	\includegraphics[width=8.5cm]{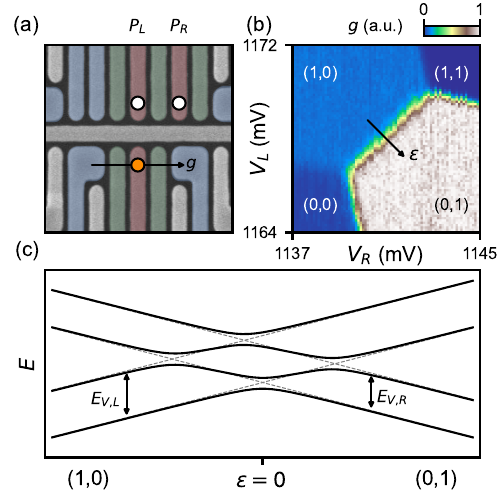}
	\caption{
    (a) False-color scanning electron microscope image of a device identical to the one used in this experiment. A DQD (white circles) is defined using plunger (red) and barrier (green) gates in the upper-half of the device. A dot charge sensor (DCS) is formed in the lower-half of the device (orange circle). Charge reservoirs are accumulated beneath the blue gates. (b) DCS conductance $g$ plotted as a function of plunger gate voltages $V_{L}$ and $V_{R}$. The detuning axis ($\varepsilon$) is overlaid on the DQD charge stability diagram. (c) DQD energy level diagram plotted as a function of $\varepsilon$ in the coupled (solid lines) and uncoupled (dashed lines) regimes. The valley splitting of the left (right) dot is denoted by $E_{V,L}$ ($E_{V,R}$).
    }
	\label{fig1}
\end{figure}

Conveyor-mode shuttling \cite{Volmer2024,Volmer2026} and DAPS \cite{Marcks2025} have recently been employed to probe variations in valley splittings across a device. However, the necessity of fabricating QD arrays with consistently large valley splittings motivates the establishment of a rapid prototyping system capable of establishing correlations between materials properties and valley splittings. In particular, spectroscopy of valley states at higher temperatures would open the door to the use of high throughput cryostats and possibly cryogenic probe stations.\cite{Neyens2024}

In this Letter, we evaluate the performance of PGS \cite{elzerman2004_PGS,yang2012,zajac2016} and DAPS \cite{chen_detuning_2021} at temperatures that can be attained in relatively inexpensive pumped helium cryostats. A valley splitting of $\sim$120 $\mu$eV is resolved using PGS up to 300 mK and DAPS up to 700 mK. From our measurements, we estimate that valley splittings as small as $\sim$210 $\mu$eV ($\sim$86 $\mu$eV) can be resolved with PGS (DAPS) at an electron temperature $T_e$ $\sim$ 700 mK. We attribute the difference in resolution to the spectral broadening mechanism associated with each measurement approach. PGS involves dot-lead transitions and is limited by thermal broadening of the Fermi reservoir. \cite{elzerman2004_PGS} On the other hand, DAPS involves interdot charge transitions, with a resolution dictated by charge dephasing rates near anti-crossings in the double quantum dot (DQD) energy level spectrum.\cite{chen_detuning_2021}

\begin{figure}[t!]
	\centering
    \includegraphics[width=8.5cm]{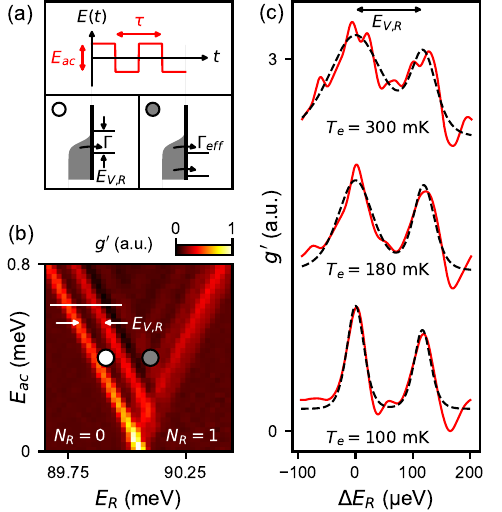}
	\caption{Temperature dependence of PGS. (a) Top panel: A square pulse repeatedly loads and unloads an electron from the QD. Bottom panel: The tunnel rate is dependent on the number of energetically accessible levels (see main text). (b) Derivative of the DCS conductance, $g'$, plotted as a function of $E_R$ and $E_{ac}$. The white and gray markers denote the configurations shown in the bottom panel of (a). (c) Line cuts acquired at the white horizontal line in (b) for three different values of $T_e$ with $E_{V,R}~\sim~120~\mu \text{eV}$. Data are smoothed and normalized (see main text), then fit to the derivative of the sum of two Fermi functions (dashed lines), yielding the valley splitting $E_{V,R}$.
    }
	\label{fig2}
\end{figure}

Measurements are performed on the Intel triple QD device shown in Fig.~\ref{fig1}(a).\cite{Neyens2024,George2024} A DQD is defined beneath plunger gates $\mathrm{P_L}$ and $\mathrm{P_R}$. A dot charge sensor (DCS) with conductance $g$ is formed in the lower portion of the device. Figure~\ref{fig1}(b) shows the DQD charge stability diagram in the vicinity of the (1,0)-(0,1) interdot charge transition. The electronic configuration of the DQD is denoted $(N_L,N_R)$, where $N_L~(N_R)$ is the number of electrons in the left (right) QD. A typical energy level spectrum of a single electron Si DQD is plotted in Fig.~\ref{fig1}(c) as a function of level detuning $\varepsilon$. \cite{Burkard2016} The valley splitting of the left (right) QD is $E_{V,L}~(E_{V,R})$. For this device, $\varepsilon = \alpha(V_R-V_L)$, where $\alpha=85$ $\mu$$ \text{eV}/\text{mV}$ is the lever arm extracted from the thermal broadening of the interdot charge transition at $T_e$ = 300 mK.\cite{DiCarlo2004}

The energy level spectrum of the right QD is next determined using PGS. A square pulse with amplitude $E_{ac}$ and period $\tau=1~\mu\text{s}$ is continuously applied to $\mathrm{P_R}$, as shown in the top panel of Fig.~\ref{fig2}(a). The square pulse repeatedly drives dot-lead transitions for the right QD. For small $E_{ac}$ only the ground state of the QD is accessible and tunneling takes place with a rate $\Gamma$. At larger $E_{ac}$, an electron can load into either the ground state or excited state, resulting in an effective tunneling rate $\Gamma_{\rm eff}$ $>$ $\Gamma$. Figure~\ref{fig2}(b) shows a PGS spectrum acquired at $T_e$~=~100~mK. Here the derivative of the DCS conductance, $g'=dg/dE_{R}$, is plotted as a function of $E_{ac}$ and $E_{R}$. Voltages are converted into energy units using the right gate lever arm $\alpha_R=78$ $\mu$$ \text{eV}/\text{mV}$ determined from finite bias triangles.\cite{johnson2005} We extract a right dot valley splitting $E_{V,R}$~=~$117\pm17$ $\mu$eV.

\begin{figure}[tb!]
	\centering
	\includegraphics[width=8.5cm]{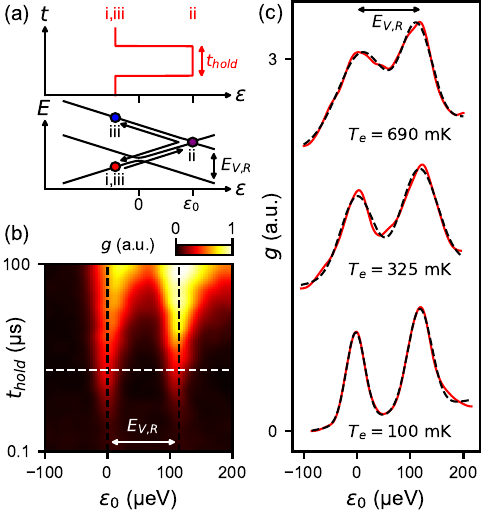}
	\caption{Temperature dependence of DAPS. (a) Top panel: Pulse sequence. Bottom panel: Energy level diagram illustrating the DAPS protocol. If point ii is near an anti-crossing, enhanced charge dephasing yields a mixed charge state.
    (b) DAPS data obtained from the right QD. $g$ is plotted as a function of $t_{hold}$ and $\varepsilon$.
    (c) Line cuts acquired at the white horizontal dashed line in (b) for three different values of $T_e$ with $E_{V,R}~\sim~120~\mu \text{eV}$. Data are smoothed and normalized (see main text), then fit to the sum of two Gaussians. The functional temperature range of DAPS exceeds PGS by nearly a factor of two.}
	\label{fig3}
\end{figure}

The high temperature performance of PGS is evaluated in Fig.~\ref{fig2}(c) by repeating the measurement with fixed $E_{ac} = 0.624$ meV and increasing $T_e$. These data are plotted as a function of $\Delta$$E_R$, where $\Delta$$E_R$ = 0 corresponds to the condition where the right QD ground state crosses the Fermi level, such that excited state energies can be directly read off the x-axis. With $T_e$ = 100 mK, the ground and excited valley state of the right dot are clearly visible. However, the PGS features broaden significantly with increasing temperature, with the peaks nearly merging around $T_e$ = 300 mK. The data are fit to the sum of the derivative of two Fermi functions and the valley splitting is extracted from the distance between the two peaks, yielding $E_{V,R}$ = $117\pm17$ $\mu$eV, $120\pm29$~$\mu$eV, and $119\pm41$~$\mu$eV in order of increasing $T_e$. The uncertainty is estimated as the average of the half-width-at-half maximum of the derivative of the two Fermi functions. The $T_e$ extracted from the fits are consistent with the thermal broadening of the ground state transition with no square wave applied.\cite{Zajac2016_2} The data in Fig.~\ref{fig2} are smoothed using a Gaussian filter with a standard deviation value of 8.2~$\mu$eV, which is below the thermal energy $k_BT_e$ at $T_e$ = 100 mK, where $k_B$ = 86 $\mu$eV/K is Boltzmann's constant. The data in Fig.~\ref{fig2} are normalized by the maximum value of $g'$ in each dataset.

Energy level spectroscopy of the right QD is independently performed with DAPS [Fig.~\ref{fig3}]. \cite{chen_detuning_2021} The DAPS pulse sequence initializes an electron in the ground state of the left QD at negative detuning [position i in Fig.~\ref{fig3}(a)]. The detuning is then non-adiabatically increased to a value $\varepsilon_0$ [position ii in Fig.~\ref{fig3}(a)] and held there for a time $t_{hold}$. The return charge state is measured for 10 $\mu$s after non-adiabatically pulsing back to position iii. If $\varepsilon_0$ coincides with a level anti-crossing, the right dot ground state will rapidly mix with left dot charge states, yielding some (0,1) charge state occupation.

Figure~\ref{fig3}(b) shows DAPS data for the right QD. Here $g$ is plotted as a function of $t_{\rm hold}$ and $\varepsilon_0$ at $T_e$~=~100~mK. The DAPS feature around $\varepsilon$ = 0 corresponds to the left dot ground state -- right dot ground state anti-crossing, while the feature around 120 $\mu$eV corresponds to the anti-crossing between the left dot ground state and right dot excited valley state. From these measurements, we obtain $E_{V,R}$ = 120 $\pm$ 24 $\mu$eV. 

The high temperature performance of DAPS is investigated in Fig.~\ref{fig3}(c) with fixed $t_{hold}$ = 2 $\mu$s. In contrast with the PGS data, the valley splitting is clearly resolvable at $T_e$ = 325 mK, and even at $T_e$ = 690 mK. Each of the three data sets is fit to the sum of two Gaussians and the valley splitting is defined as the peak separation. The uncertainty is estimated as the average of the half-width-at-half maximum of the two Gaussians. $E_{V,R}$ = 120 $\pm$ 24~$\mu$eV, 117 $\pm$ 38~$\mu$eV, and 104 $\pm$ 43~$\mu$eV in order of increasing $T_e$. Similar to Fig.~\ref{fig2}, the data in Fig.~\ref{fig3} are smoothed using a Gaussian filter with a standard deviation value of 8.2 $\mu$eV and are normalized by the maximum value of $g$ in each dataset.

\begin{figure}[tb!]
	\centering
	\includegraphics[width=8.5cm]{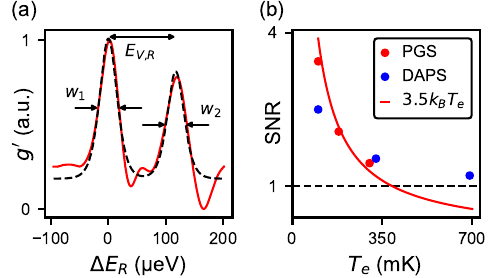}
	\caption{
    (a) Sample PGS valley splitting measurement depicting the variables used to calculate the SNR with $E_{V,R}~\sim~120~\mu \text{eV}$.
    (b) SNR calculated from $E_{V,R}$, $w_1$, and $w_2$ extracted from Figs.~\ref{fig2}(c) and ~\ref{fig3}(c). The red line shows the theoretical linewidth of $3.5k_BT_e$ for PGS. For $E_{V,R}$ = 120~$\mu$eV, DAPS provides an SNR $>$ 1 at $T_e$ $\sim$ 700~mK.
    }
	\label{fig4}
\end{figure}

The broadening of the DAPS data is significantly smaller than the PGS data. We attribute the difference in resolution to the contrasting electron transitions involved in the two measurements. PGS involves dot-lead transitions, where QD electrons are coupled to thermally broadened charge reservoirs with a linewidth of $3.5k_BT_e$.\cite{mills2022high,Yoo2026} However, DAPS only involves interdot charge transitions between zero-dimensional electronic states. Spectral broadening of the DAPS data is therefore dominated by charge dephasing at the anti-crossings.\cite{chen_detuning_2021,duan2026} 

We quantify the resolution of the spectroscopy techniques by defining a signal-to-noise ratio (SNR) as SNR $=E_{V,R}/\overline{w}$, with $\overline{w}=(w_1+w_2)/2$, where $w_1~(w_2)$ is the full-width-at-half-maximum of the ground (excited) valley peak, as illustrated with an example data set in Fig.~\ref{fig4}(a). The data points in Fig.~\ref{fig4}(b) show the SNR extracted from the PGS data in Fig.~\ref{fig2}(c) and DAPS data in Fig.~\ref{fig3}(c). The red line is the expected PGS SNR scaling with temperature due to 3.5$k_BT_e$ broadening of the Fermi function. At very low $T_e$, PGS results in higher SNR than DAPS. However, as $T_e$ increases, the PGS SNR rapidly degrades due to direct coupling with a Fermi sea, while charge dephasing is only modestly enhanced. As a result, DAPS yields an SNR $\approx$ 1 at an electron temperature that is twice as high as PGS, making DAPS the favored measurement approach in high throughput screening cryostats.

Lastly, we estimate the minimum valley splitting that can be resolved at $T_e$ $\sim$ 700 mK for each approach. As SNR = 1 denotes the limit of resolving a valley splitting, $E_{V,R} = \overline{w}$ represents the minimum valley splitting that can resolved for a given $\overline{w}$, set by $T_e$ and the valley spectroscopy method used. For PGS, $T_e$ $\sim$ 700 mK results in $\overline{w}$ $\sim$ 210 $\mu$eV, estimated from the $3.5k_BT_e$ broadening identified in Fig.~\ref{fig2}(c). Therefore, the minimum valley splitting that can be resolved using PGS is $E_{V,R}$ $\sim$ 210~$\mu$eV. For DAPS, the energy resolution at $T_e$ $\sim$ 700~mK is superior to PGS, with a minimum valley splitting of $\sim$86 $\mu$eV, directly extracted from $\overline{w}$ in Fig.~\ref{fig3}(c).

In conclusion, we characterize the performance of two common energy level spectroscopy techniques at elevated temperatures. The resolution of PGS is limited by the thermal broadening of the Fermi function. In contrast, the spectroscopically relevant portion of the DAPS pulse sequence does not involve coupling to a Fermi sea and provides signatures of valley states that persist up to $\sim$700 mK. Due to its superior energy resolution, we propose the use of DAPS to extract valley splittings in high throughput screening cryostats. The large-scale characterization of QD arrays will provide crucial feedback on the design of Si/SiGe devices that consistently yield large valley splittings.

\begin{acknowledgments}
We acknowledge the support of AFOSR grant FA9550-23-1-0710 and ARO grants W911NF-23-1-0104 and W911NF-22-2-0037. The views and conclusions contained in this document are those of the authors and should not be interpreted as representing the official policies, either expressed or implied, of the Army Research Office or the U.S. Government. The U.S. Government is authorized to reproduce and distribute reprints for Government purposes notwithstanding any copyright notation herein. We acknowledge support from Intel Corporation for providing the device.
\end{acknowledgments}

\bibliography{RMP_master2_bib_v7}

\end{document}